# Centrifugal Microfluidics for Biomedical Applications

*Yoon-Kyoung Cho - Ulsan National Institute of Science and Technology (UNIST)*

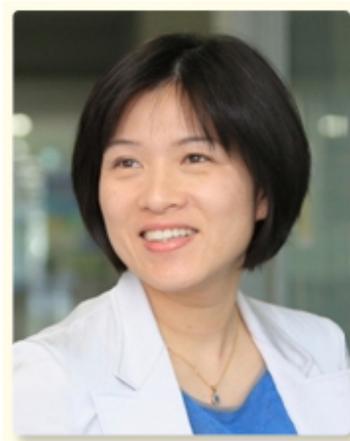

## Biography
***Yoon-Kyoung Cho*** *is currently a full professor in Biomedical Engineering at UNIST and a group leader in the Center for Soft and Living Matter at the Institute for Basic Science (IBS), Republic of Korea. She received her Ph.D. in Materials Science and Engineering from the University of Illinois at Urbana-Champaign in 1999, having obtained her M.S. and B.S. in Chemical Engineering from POSTECH in 1994 and 1992, respectively. She worked as a senior researcher (1999–2008) at Samsung Advanced Institute of Technology (SAIT), where she participated in the development of in vitro diagnostic devices for biomedical applications. Since she joined UNIST in 2008, she has been the chair of the school of Nano-Bioscience and Chemical Engineering (2008–2014) and the school of Life Sciences (2014–2015) and the director of World Class University (2009–2013) and BK21 (2013–2015) programs. She named a Fellow of the Royal Society of Chemistry in 2016. Her research interests range from basic sciences to translational research in microfluidics and nanomedicine. Current research topics include a lab-on-a-disc for the detection of rare cells and extracellular biomarkers, quantitative analysis of single cells, and system analysis of cellular communication. http://fruits.unist.ac.kr*

## Abstract
The development of a rapid, miniaturized, and efficient on-chip sample preparation for "real" sample analysis remains a major bottleneck for the realization of a lab-on-a-chip approach in point-of-care diagnostics. We developed a fully integrated and automated lab-on-a-disc using centrifugal microfluidics to provide a "sample-in and answer-out" type of biochemical analysis solution with simple, size-reduced, and cost-efficient instrumentation.[1] Here, I present various examples of the fully integrated "lab-on-a-disc" developed for broad applications ranging from medical diagnostics to food, environment, and energy applications **(Fig. 1A).**

## Active valves on a spinning disc
We pioneered the concept of laser-irradiated ferrowax microvalves (LIFM) with colleagues at Samsung Advanced Institute of Technology (SAIT), which provided a simple and robust tool for obtaining fluidic control on a spinning disc.[2] The key achievement of this work was the rapid and wireless actuation of multiple valves by simple laser irradiation on nanoheaters, which are 10-nm-sized ferro-oxide nanoparticles dispersed in paraffin wax **(Fig. 1B).**[2] The response time of both the normally open and normally closed valves was very short, and the actuation of the valves was independent of the sequence of the spin speed, sample type, or material properties of the substrates. More recently, we demonstrated the temperature-stable and reversible actuation of individually addressable diaphragm valves, which can be integrated on a disc for broader applications.[3] I believe that such integration of centrifugal pumping with remote control of opening and closing the valves provides an ideal combination to realize a fully integrated lab-on-a-chip system for processing real samples. I will here provide a few representative examples to demonstrate this concept.

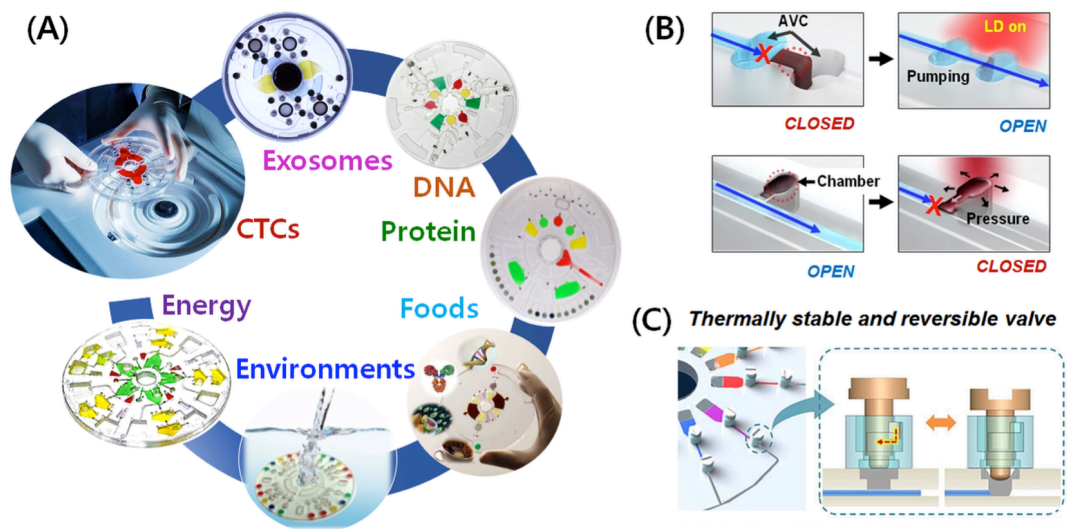

Figure 1. (A) Examples of the fully integrated lab-on-a-chip for biomedical[4, 6-13], food[5, 14-15], environmental[16], and energy[17] applications. (B) Working principles of normally open (top) and normally closed (bottom) laser-irradiated ferrowax microvalves2. (C) Thermally stable and reversible diaphragm valves.[3]

## Fully integrated DNA analysis on a disc

My first attempt to build a fully integrated lab-on-a-disc involved the isolation of pathogen-specific DNA starting from 100 μL of whole blood spiked with hepatitis B virus.[4] The entire process of plasma separation from whole blood, virus enrichment using magnetic beads modified with pathogen-specific antibodies, washing, and viral DNA extraction could be completed within 12 min in a fully automated manner. The viral DNA yield quantified by real-time polymerase chain reaction (PCR) was as good as that achieved using a commercially available product. In this case, we used a diode laser for dual purposes: for valve actuation and laser-irradiated viral DNA extraction. More recently, we further developed the system so that the same diode laser could be used for local heating to perform on-disc PCR or isothermal amplification.[5]

## Fully automated enzyme-linked immunosorbent assay (ELISA) starting from whole blood

Although many microfluidic devices have been reported to date, very few are sufficiently robust, automated, and user-friendly for practical usage. Some of the major challenges derive from the complex nature of the biological samples and the difficulties in developing an effective human-chip interface. I believe that microfluidic technologies must be simple-to-use to allow a non-expert to readily employ the technology in routine tests. In this regards, we developed a fully automated, beads-based ELISA for plasma protein analysis starting from whole blood, which can be accomplished within 30 min.[6] In addition, both the immunoassay and blood chemistry analysis of multiple analytes starting from whole blood were fully automated on a self-contained disc.[7] The unique actuation of LIFM enabled two independent assay protocols to be integrated and simultaneously performed starting from whole blood samples. Subsequently, the sensitivity of multiplex protein detection[8] was further enhanced by utilizing flow-enhanced electrochemical detection[9] or $TiO_2$ nanofiber arrays.[10]

## Isolation of circulating tumor cells (CTCs) from whole blood

Liquid biopsy has great potential for the detection of cancer metastasis and in the real-time monitoring of cancer treatments. Among the existing technologies, size-based isolation methods provide antibody-independent, relatively simple, and high-throughput protocols. However, the clogging issues and lower than desired recovery rates and purity are the key challenges. To overcome these limitations, we used a tangential flow filtration system integrated on a disc, where the centrifugal force and filtration directions are in a perpendicular direction. In addition, the fluidic resistance of the inlet and outlet of the filter membrane was adjusted to provide a continuously wetted membrane for gentle yet efficient size-based filtration. We achieved highly sensitive (95.9 ± 3.1% recovery rate), selective (>2.5 log depletion of white blood cells), rapid (>3 mL/min), and label-free isolation of viable CTCs from whole blood without prior sample treatment using a stand-alone lab-on-a-disc system equipped with fluid-assisted separation technology (FAST).[11-12] Not only the number of CTCs but also the mutation analysis from isolated CTCs from whole blood samples taken from patients with lung and stomach cancers validated the clinical utility of CTCs analysis as a minimally invasive liquid biopsy tool.

## Fully automated extracellular vesicle (EV) isolation from urine samples

Next, we developed a rapid, label-free, and highly sensitive method for the isolation and quantification of EVs using a lab-on-a-disc integrated with two nanofilters (Exodisc).[13] Urinary EVs from patients with bladder cancer could be automatically enriched within 30 min using a tabletop-sized centrifugal microfluidic system, which is the same instrument used for the CTC isolation described above. Compared with the gold-standard ultracentrifugation method, the concentration of the mRNA retrieved by the disc system was increased by >100-fold. We believe that this revolutionary CTCs and EV isolation method can potentially contribute to expediting the acceptance of liquid biopsy-based cancer diagnostics as a standard practice in clinical settings.

## Lab-on-a-disc for food, energy, and environmental applications

Centrifugal microfluidics is a platform technology that can be utilized not only for biomedical diagnostics but also for much broader areas including food, energy, and environmental applications. Through collaboration with experts in various disciplines, we demonstrated the possibility of point-of-care diagnostics in non-medical applications. For example, the molecular detection of food-borne pathogens, e.g., *Salmonella* from milk samples, could be completed within 30 min in a fully automated manner. Three main functions—DNA extraction, isothermal recombinase polymerase amplification, and color detection—were integrated on a disc.[5] We also demonstrated the ability of lab-on-a-disc to detect the caffeine concentration from various types of beverage samples. The caffeine content was displayed like a traffic light, with green and red color representing low and high caffeine contents, respectively, using a novel aqueous-phase fluorescent caffeine sensor named "Caffeine Orange".[14] In addition, the total phenolic content and antioxidant activity of beverages such as fruit juices, tea, wine, and beer could be measured by the lab-on-a-disc.[15] For environmental monitoring, we developed a lab-on-a-disc for the simultaneous determination of five major inorganic nutrients related to harmful algal blooms from ocean water.[16] With respect to bioenergy application, we demonstrated the utility of a lab-on-a-disc for the rapid on-site quantification of lipids from microalgal samples. The fully automated serial process, involving cell sedimentation and lysis, liquid-liquid extraction of lipids, and colorimetric detection of lipid contents, could be performed on a disposable, organic solvent-tolerable (for n-hexane, ethanol) plastic disc fabricated using thermal fusion bonding and carbon dot-based valving techniques.[17]

## Coda

In summary, we have demonstrated that spinning discs integrated with active control of microvalves could provide an extremely robust and versatile platform technology that can be used for the processing of real samples, with a miniaturized format that can deliver more accurate outputs in a much simpler, faster, and cheaper manner than conventional technologies. Among the many applications possible and demonstrated, our major interests currently lie in the potential for personalized medicine. We believe that this technology will drastically enhance the quality of life for an aging population and reduce the cost of healthcare through facilitating screenings of the genetic make-up at the individual level. For example, it will be a dream tool if it allows for the early diagnosis of cancer metastasis together with the quantification and characterization of the genetic information of cancer malignancy, thereby offering personalized therapy while achieving 'real-time' monitoring to continuously check how the cells are responding to the treatment. Recent advances in microfluidics have provided evidence that rare cells and molecules such as CTCs, exosomes, and cell-free DNAs could serve as non-invasive yet reliable diagnostic tools and prognostic biomarkers. However, the major hurdle remains the extreme rarity of target cells and molecules that exist in a large background population. Conventional techniques have limitations in isolation efficiencies and specificity as well as significant sample loss and long processing times for such fragile cells.

We have recently developed a lab-on-a-disc for size-based isolation of CTC and EVs and tested it with blood and urine samples from patients with various types of cancer. As briefly introduced above, these tools are simple to use, gentle yet fast, and the isolation sensitivity and selectivity are good, thereby allowing for real samples such as whole blood or urine to be used directly without prior treatment. We plan to use the filtered rare cells or EVs for follow-up molecular analysis and other functional analyses. Although it is important to further develop the technology to realize its application as a novel cancer metastasis diagnostic device, I believe it will also serve as an essential tool to provide rare cells and EVs originating from cancer patients rather than derived from a cell line. Obtaining such resources will facilitate the discovery of biochemical and physical information at the molecular level and help to expand our knowledge of basic biology and medicine. This will require development of a single-cell isolation method as well as genomic, proteomic, physical, and functional analysis technologies, and rapid information processing capabilities for systems analysis to handle the large amounts of data.

**Acknowledgement**

The work presented in this abstract would not have been possible without the team effort of graduate students, researchers, and numerous colleagues and collaborators at SAIT, UNIST, IBS, Clinomics, Pusan National University Hospitals, and many more. I express my gratitude to all of the patients and healthy volunteers who contributed biological samples for this study.